\documentclass[a4paper,12pt]{article}
\usepackage{amsmath}
\usepackage{amssymb}
\usepackage{amsfonts}

\usepackage{hyperref}
\usepackage{mathrsfs}
\usepackage{color}
\usepackage{cancel}
\usepackage{graphicx}

\def\beq{\begin{equation}}
\def\eeq{\end{equation}}
\def\bea{\begin{eqnarray}}
\def\eea{\end{eqnarray}}
\newcommand{\ba}{\begin{array}}
\newcommand{\ea}{\end{array}}

\def\<{\left\langle}
\def\>{\right\rangle}

\begin{document}

\bibliographystyle{OurBibTeX}

\begin{titlepage}
\begin{center}
{
\sffamily
\Large
 \bf
Little $Z'$ Models
}
\\[12mm]
Alexander Belyaev\footnote{E-mail: \texttt{a.belyaev@soton.ac.uk}},
S.~F.~King\footnote{E-mail: \texttt{sfk@hep.phys.soton.ac.uk}},
Patrik Svantesson\footnote{E-mail: \texttt{p.svantesson@soton.ac.uk}}
\\[1mm]
{\small\it
School of Physics and Astronomy,
University of Southampton,\\
Southampton, SO17 1BJ, U.K.
}
\end{center}
\vspace*{1.00cm}

\begin{abstract}
\noindent

{We propose a new class of models called {\it Little $Z'$} models in order
to reduce the fine-tuning due to the current experimental limits on the $Z'$ mass in
$E_6$ inspired supersymmetric models, where the Higgs doublets are charged under the extra $U(1)^\prime$ gauge group. The proposed {\it Little $Z'$} models allow a lower mass $Z'$ due to the spontaneously broken extra $U(1)'$ gauge group having a reduced gauge coupling.}
We show that reducing the value of the extra gauge coupling relaxes the experimental limits, leading to the possibility of low mass $Z'$ resonances, for example down to 200 GeV, which may yet appear in LHC searches. 
Although the source of tree level fine-tuning due to the $Z'$ mass is reduced in Little $Z'$ models, it typically does so at the expense of increasing the{ vacuum expectation value of the $U(1)^\prime$-breaking standard model singlet field}, {reducing the fine-tuning to similar 
levels} to that in the Minimal Supersymmetric Standard Model.
\end{abstract}

\end{titlepage}
\newpage
\setcounter{footnote}{0}

\section{Introduction}

The Large Hadron Collider (LHC) has so far not seen any signal of
new physics beyond the standard model (BSM).
On the other hand ATLAS and CMS have recently observed a new state
consistent with a Standard-Model-like 
Higgs boson at $m_h = 125-126$
GeV \cite{:2012gk,:2012gu}, which is within the range for it to be
consistent with the lightest Higgs in supersymmetric models.  
In the minimal supersymmetric standard model (MSSM) the light Higgs mass
at tree-level is bounded from above by the $Z$ boson mass ($M_Z$).
The large radiative contributions from stops needed to raise it to the
observed value typically imply very large fine-tuning. 

Conventional $E_6$ inspired SUSY models
involve both a singlet generated $\mu$ term, denoted $ \mu_{\text{eff}}$, and 
a massive $Z'$ gauge boson at the TeV scale. 
Such models can increase the tree level physical Higgs boson mass above the $M_Z$
limit of the MSSM, due to both F-term contributions of the singlet and the D-term contributions
associated with the $Z'$, allowing lighter stop masses and hence reducing fine-tuning due to stop loops.
The exceptional supersymmetric standard model (E$_6$SSM) \cite{King:2005jy, King:2005my}
is an example of such a model, inspired by the $E_6$ group.
It involves an extra singlet responsible for $ \mu_{\text{eff}}$ and an extra 
$U(1)$ gauge symmetry at low energy, giving both new F-term and D-term contributions at tree level to the 
light Higgs mass, which is larger than both the MSSM and the next-to-minimal supersymmetric standard model (NMSSM)\cite{genNMSSM2}.
In the E$_6$SSM the light Higgs mass is given by,
\begin{equation}  \label{mh}
m_h^2 \approx  M_Z^2 \cos^2 2\beta +  \frac{ \lambda^2}{2} v^2 \sin^2 2 \beta  +  g_1'^2 v^2 (Q_1 \cos^2 \beta + Q_2 \sin^2 \beta)^2 + \Delta m_h^2.
\end{equation}
where $\tan\beta$ is the ratio between the two Higgs doublets' vacuum expectation values (VEVs), $\lambda$ is the Yukawa coupling of 
the singlet field to the Higgs doublets, the extra $U(1)'$ gauge group has a gauge coupling $g_1'$
and $\Delta m_h^2$ represents loop corrections.

Eq.\ref{mh} exhibits two extra terms proportional to $v^2$, relative to the MSSM,
which contribute at tree level to the Higgs mass squared.
This means that the E$_6$SSM
permits lower stop masses than in the MSSM (or the NMSSM) corresponding to 
lower values required for the radiative correction term $\Delta m_h^2$.
However, as we shall discuss, one of the minimisation conditions
of the E$_6$SSM can be written in the form,
\beq
\label{mz0}
c \frac{M_Z^2}{2}=-\mu_{\text{eff}}^2+  \frac{ (m_d^2 - m_u^2 \tan^2\beta) }{ \tan^2\beta - 1 } +d \frac{M_{Z'}^2}{2},
\eeq
where $c,d$ are functions of $\tan \beta$ which are of order $\sim \mathcal{O}(1)$,
$m_d^2, m_u^2$ are soft Higgs mass squared parameters, $M_{Z'}\sim g_1's$
and  $ \mu_{\text{eff}} \sim \lambda s $ arise from the singlet VEV $s$.
Written in this form it is clear that there is a new source of tree-level fine-tuning, due to the $Z'$ mass squared term in Eq.\ref{mz0}, which 
will increases quadratically as $M_{Z'}^2$, eventually coming to dominate the fine-tuning for large enough 
values of $M_{Z'}$.
This tree-level fine-tuning can be compared to that due to $\mu_{\text{eff}}$ which typically requires this
quantity to be not much more than 200 GeV, and similar limits also apply to $M_{Z'}$.
With the current CMS experimental mass limit for the $Z'$ in the E$_6$SSM of 
$M_{Z'}\gtrsim 2.08$ TeV \cite{Chatrchyan:2012it} it is clear that there is 
already a significant, perhaps dominant, 
amount of fine-tuning due to the $Z'$ mass limit, and furthermore this source of fine-tuning
increasing quadratically with $M_{Z'}$ will rapidly overtake the logarithmic fine-tuning due to the stop mass limits,
as the experimental mass limits of both types of particles increases in the future.
This was first pointed out in \cite{Hall:2012mx} and has been discussed 
quantitatively \cite{Athron:2013ipa} in the framework of the constrained E$_6$SSM \cite{Athron:2011wu},
where it has been verified that this new source of fine-tuning dominates over all other sources.

In this paper we propose a new class of models called Little $Z'$ models which differ from the usual 
class of $E_6$ models by having a reduced gauge coupling $g_1'$ 
leading to the possibility of lower mass $Z'$ bosons.
Such a reduction in the gauge coupling $g_1'$ at the unification scale 
has some motivation from F-theory constructions \cite{Callaghan:2012rv}.
We show that reducing $g_1'$ relaxes the experimental limit on the $Z'$ mass,
allowing a lighter value and hence reducing the tree-level fine-tuning
associated with $E_6$ models. We show that, although 
for sufficiently small values of $g_1'$ the new source of fine-tuning due to 
the $Z'$ mass can be essentially eliminated, it does so 
at the expense of increasing the singlet vacuum expectation value, leading to overall fine-tuning similar to that in the Minimal Supersymmetric Standard Model. We emphasise that the main prediction of Little $Z'$ models
is the presence of weakly coupled low mass $Z'$ resonances, perhaps as low as 200 GeV.

The layout of the remainder of the paper is as follows.
In section 2 we briefly review the E$_6$SSM, followed by a discussion of the Electroweak Symmetry Breaking (EWSB) conditions and the impact of the $Z'$ mass on fine-tuning in section 3. Little $Z'$ models are introduced in section 4, where the experimental limits on such a boson are studied as a function of its mass and (reduced) gauge coupling. Section 5 concludes the paper.

\section{The E$_6$SSM}

At low energies, the group structure of the 
Exceptional Supersymmetric Standard Model (E$_6$SSM) 
 is that of the Standard Model (SM), along with the additional $U(1)_N$ symmetry,

\begin{align}
E_6 &\rightarrow SU(5) \times U(1)_N \\
SU(5) &\rightarrow SU(3)_c \times SU(2)_w \times U(1)_Y 
\end{align}
The matter content of the model is contained in the complete 27-dimensional representation which decomposes under $SU(5) \times U(1)_N$ to,

\begin{equation}
27 \longrightarrow (10, 1)_i + (5^*, 2)_i + (5^*, -3)_i + (5, -2)_i + (1, 5)_i + (1, 0)_i
\end{equation}
Ordinary Quarks and Leptons are contained in the representations: $(10,1)$ and $(5^*,2).$ The Higgs doublets and exotic quarks are contained in $(5^*, -3)$ and $(5,-2).$ 
The singlets are contained in $(1,5)$, and finally the right handed neutrinos are included in $(1,0).$

Moreover, the model requires three 27 representations, hence $i =
1,2,3$, in order to ensure anomaly cancellation.  This means that
there are three copies of each field present in the model. However,
only the third generation (by choice) of the two Higgs doublets, and
the SM singlet acquire VEVs. The other two generations are called:
inert. Furthermore, in order to keep gauge coupling unification,
non-Higgs fields that come from extra incomplete $27', \bar{27}'$
representations are added to the model. As a result, a $\mu^{\prime}$
term, which is not necessary related to the weak scale, is present in
the model.

The full superpotential consistent with the low energy gauge
structure of the E$_6$SSM contains includes both $E_6$ invariant
invariant terms and $E_6$ breaking terms, full details of which are
given in \cite{King:2005jy}.

To prevent proton decay and flavour changing neutral currents a discrete
$Z_2^H$ symmetry is imposed. All superfields except the third generation
Higgs doublets and singlet are odd under this symmetry.
 The $Z_2^H$ invariant superpotential then reads,

\begin{eqnarray}
W_{\rm E_6SSM} &\approx &
\lambda_i \hat{S}(\hat{H}^d_{i}\hat{H}^u_{i})+\kappa_i \hat{S}(\hat{D}_i\hat{\overline{D}}_i)+
f_{\alpha\beta}\hat{S}_{\alpha}(\hat{H}_d \hat{H}^u_{\beta})+ 
\tilde{f}_{\alpha\beta}\hat{S}_{\alpha}(\hat{H}^d_{\beta}\hat{H}_u) \nonumber\\[2mm]
&&+\dfrac{1}{2}M_{ij}\hat{N}^c_i\hat{N}^c_j+\mu'(\hat{H}'\hat{\overline{H'}})+
h^{E}_{4j}(\hat{H}_d \hat{H}')\hat{e}^c_j+h_{4j}^N (\hat{H}_{u} \hat{H}')\hat{N}_j^c 
\nonumber \\[2mm]
&& + W_{\rm{MSSM}}(\mu=0),
\label{Eq:SupPot}
\end{eqnarray}

\noindent where the indices $\alpha, \beta = 1,2$ and $i = 1,2,3$ denote the generations. $S$ is the SM singlet field, $H_u,$ and $H_d$ are the Higgs doublet fields 
corresponding to the up and down types.  Exotic quarks and the additional non-Higgs fields are denoted by $D$ and $H'$ respectively. 

Finally to ensure that only third generation Higgs like fields get
VEVs a certain hierarchy between the Yukawa couplings must
exist. Defining $\lambda \equiv \lambda_3$, we impose
$\kappa_i\sim\lambda_3\gtrsim\lambda_{1,2}\gg
f_{\alpha\beta},\,\tilde{f}_{\alpha\beta},\,h^{E}_{4j},\,h_{4j}^N$.

\section{The Higgs potential and the EWSB conditions}

The scalar Higgs potential is,
\begin{equation}
\begin{split}
V(H_d, H_u, S) = & \ { \lambda^2 |S|^2 (|H_d|^2 + |H_u|^2 ) + \lambda^2 | H_d . Hu |^2 }\\
& + { \frac{ g_2^2 }{ 8 } (H^\dagger_d \sigma_a H_d + H^{\dagger}_u \sigma_a H_u ) (H^ {\dagger}_d \sigma_a H_d + H^{\dagger}_u \sigma_a H_u ) }\\
& + { \frac{ {g'}^2 }{ 8 } (|H_d|^2 - |H_u|^2 )^2 + \frac{ {g'}_1^2 }{ 2 } (Q_1 |H_d|^2 + Q_2 |H_u|^2 + Q_s |S|^2)^2 }\\
& + { m_s^2 |S|^2 + m_d^2 |H_d|^2 + m_u^2 |H_u|^2 }\\
& + { [\lambda A_{\lambda} S H_d . H_u + c.c. ] + \Delta_{\text{Loops}} }
\end{split}
\end{equation}
where, $g_2$, $g' (= \sqrt{3/5} g_1)$, and $g_1^\prime$ are the gauge couplings of $SU(2)_L, U(1)_Y$ (GUT normalized), 
and the additional $U(1)_N$, respectively. $Q_1=-3/\sqrt{40}, Q_2=-2/\sqrt{40},$ and $Q_s=5/\sqrt{40}$ are effective $U(1)_N$ charges of $H_u, H_d$ and $S$, respectively. 
$m_s$ is the mass of the singlet field, and $m_{u,d} \equiv m_{H_{u,d}}$.

The Higgs field and the SM singlet acquire VEVs at the physical minimum of this potential,

\begin{equation}
<H_d> = \frac{1}{\sqrt{2}}   \begin{pmatrix} v_1 \\ 0 \end{pmatrix}, \ \ \  <H_u> = \frac{1}{\sqrt{2}}   \begin{pmatrix} 0 \\ v_2 \end{pmatrix},  <S> = \frac{s}{\sqrt{2}} , 
\end{equation}
\
\\
It is reasonable exploit the fact that $s \gg v$, which will help in simplifying our master formula 
for fine-tuning as will be seen in Section 4. Then, from the minimisation conditions, 

\begin{equation}
 \frac{ {\partial V_{E_6SSM} } }{ \partial v_1 } = \frac{ \partial V_{E_6SSM} }{ \partial v_2 } = \frac{ \partial V_{E_6SSM} }{ \partial s } = 0 ,
\end{equation}
the Electroweak Symmetry Breaking (EWSB) conditions are,

\begin{equation} \label{mz}
\frac{M_Z^2}{2} = -\frac{1}{2} \lambda^2 s^2 +  \frac{ (m_d^2 - m_u^2 \tan^2\beta) }{ \tan^2\beta - 1 } + \frac{ g_1'^2 }{2} \left(Q_1 v_1^2 + Q_2 v_2^2 + Q_s s^2\right) \frac{ (Q_1 - Q_2 \tan^2\beta) }{ \tan^2\beta - 1 } 
\end{equation}

\begin{equation} \label{sin2b}
\sin 2\beta \approx \frac{ \sqrt{2} \lambda A_{\lambda} s }{ m_d^2 + m_u^2 + \lambda^2 s^2 + \frac{ g_1'^2 }{2} Q_s s^2 ( Q_1 + Q_2 ) },
\end{equation}

\begin{equation} \label{ms}
m_s^2 \approx -\frac{1}{2}  g_1'^2 Q_s^2 s^2 = - \frac{1}{2} M_{Z'}^2,
\end{equation}
where $M_Z^2=\frac{1}{4}({g'}^2+g_2^2)(v_2^2+v_1^2)$ and 
$M_{Z'}^2 \approx  g_1'^2 Q_s^2s^2$.

Eq.\ref{mz} can be written,
\begin{equation}
	\frac{M_Z^2}{2}\left( 1-\frac{g_1^{\prime 2}}{g^{\prime 2}+g_2^2}P(\tan\beta)R(\tan\beta)\right)=-\left(\frac{\lambda s}{\sqrt 2}\right)^2 + \frac{m_d^2-m_u^2\tan^2\beta }{\tan^2\beta-1}+\frac{M_{Z'}^2}{2}R(\tan\beta)
	\label{mssmcomp}
\end{equation}
where 
\begin{equation}
	R(\tan\beta)=\frac{{Q_1}-\tan^2\beta{Q_2}}{\tan^2\beta-1}
\end{equation}
and  
\begin{equation}
	P(\tan\beta)=4\left(  \frac{Q_1(1-\frac{Q_1}{Q_S})+\tan^2\beta Q_2(1-\frac{Q_2}{Q_S})}{\tan^2\beta+1}\right)
\end{equation}
If one takes $g_1^{\prime}=0$ we have $M_{Z^{\prime}}=0$ and the factor in front of $M_Z^2$ in \eqref{mssmcomp} is equal to one and we recover the well known MSSM relation between $M_{Z}$, $\mu(=s\lambda/\sqrt{2})$ and the soft Higgs masses $m_1$, $m_2$. 
Written in this form, which may be compared to Eq.\ref{mz0} but with the coefficients $c,d$ explicitly given, it is clear that fine-tuning 
will increase quadratically as $M_{Z'}$ increases.

\begin{figure}[h!]
	\begin{center}
		\includegraphics[width=.6\linewidth]{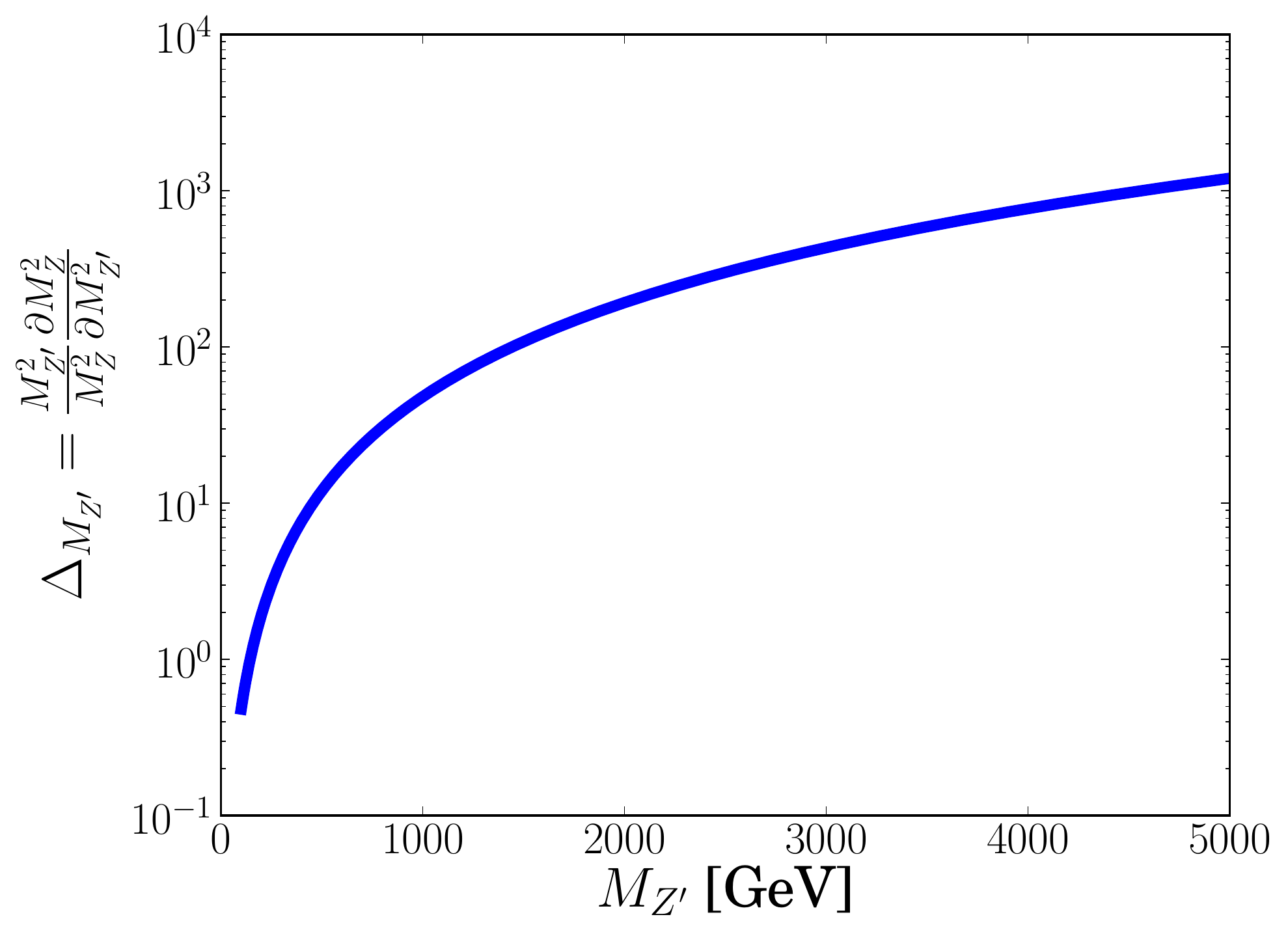}
	\end{center}
	\caption{Contribution to fine-tuning from the $Z^{\prime}$ mass.}
	\label{fig:finetuning}
\end{figure}

To avoid any fine-tuning we would like to keep $\mu \sim M_{Z'} \sim 200$ GeV or less.
This motivates the main idea of this paper, namely to relax the CMS experimental
mass limit of $M_{Z'}\gtrsim 2.08$ TeV \cite{Chatrchyan:2012it} down to $M_{Z'} \sim 200$ GeV
by reducing its gauge coupling $g_1'$.
Indeed, as we shall see, 
such a low value of $M_{Z'} \sim 200$ GeV may be made consistent with the experimental limit 
by choosing $g_1'\sim 10^{-2}\times 0.46$ and $s\sim 20\times 2.75\sim 55$ TeV.
In order to keep $\mu $  close to the electroweak scale this requires a very small value of 
$\lambda \sim g_1'$.
In Fig.~\ref{fig:finetuning} the contribution $\Delta_{M_{Z'}}$ to fine-tuning 
from $M_{Z^{\prime}}$ is plotted, where  $\Delta_{M_{Z'}}$ is defined as follows.
\begin{equation}
\Delta_{M_{Z'}}= \frac{M_{Z'}^2}{M_Z^2} \frac{\partial M_Z^2}{\partial M_{Z'}^2}
\label{eq:fine-tuning}
\end{equation}

We emphasise that the appearance of $M_{Z'}$ in the tree-level minimisation condition is characteristic
of all SUSY $Z'$ models where the usual Higgs doublets carry $U(1)'$ charges (e.g. it applies to all $E_6$ models
but not, for example the $U(1)_{B-L}$ model.)
This provides a motivation for Little $Z'$ models in which the extra gauge coupling $g_1'$ is reduced and 
the experimental lower bound on $M_{Z'}$ may be relaxed.

\section{Little $Z'$ Models}
In general Little $Z'$ Models can be defined by the gauge group
\beq
SU(3)_c\times SU(2)_L\times U(1)_Y \times U (1)^\prime
\label{low}
\eeq
where the Standard Model is augmented by an additional $U (1)^\prime$ gauge 
group with a gauge coupling $g_1'$ which is significantly smaller than the hypercharge gauge coupling
$g'$. The $U(1)^\prime$ gauge group is broken at low energies
giving rise to a massive $Z^\prime$ gauge boson with couplings to a SM 
fermion $f$ given by \cite{Accomando:2010fz}:
$$
{\mathcal L}_{NC}=\frac{g_1'}{2}Z^\prime_{\mu}\bar{f}\gamma^\mu (g_V^f  - g_A^f\gamma^5) f .
$$
The values of $g_V^f,g_A^f$ depend on the particular choice of $U (1)^\prime$
and on the particular fermion $f$. We assume universality amongst the three 
families. More explicitly, this assumption implies that
$g_V^u=g_V^c=g_V^t$, $g_V^d=g_V^s=g_V^b$, $g_V^e=g_V^{\mu}=g_V^{\tau}$, 
and $g_V^{\nu_e}=g_V^{\nu_\mu}=g_V^{\nu_\tau}$ for up-quarks, down-quarks, 
charged leptons and neutrinos respectively. The axial couplings, $g_A^f$,
behave accordingly. 

In a given model there are eight model 
dependent couplings of the extra $Z^\prime$ boson to SM fermions, that is 
$g_{V,A}^f$ with $f=u,d,e,\nu_e$. These are fixed by group theory, so cannot be changed
for a given model. However the low energy $U(1)'$ gauge coupling $g_1'$ is fixed by a unification condition.
E.g. in E$_6$SSM $g_1'\approx 0.46$ which is approximately equal to the (GUT normalised) hypercharge gauge coupling.
If unification of $g_1'$ with the other gauge couplings
is relaxed, then $g_1'$ becomes a free parameter. In this paper we are interested in taking it to be
smaller than the GUT prediction, namely we shall consider $g_1' \ll g'\approx 0.46$, keeping $g_V^f,g_A^f$ fixed at their
model predictions.

Specializing to the charged lepton pair production cross-section relevant for
the first runs at the LHC, the cross-section may be written at the leading order (LO)
as \cite{Accomando:2010fz}:
\beq
\sigma_{\ell^+\ell^-}^{LO} = \frac{\pi}{48 s}
\left[ c_uw_u(s,M_{Z'}^2)+ c_dw_d(s,M_{Z'}^2)\right]
\label{eq:ll}
\eeq
where the coefficients $c_u$ and $c_d$ are given by:
\beq
c_{u}=\frac{{g_1'}^2}{2}({g_V^u}^2+{g_A^u}^2)Br(\ell^+\ell^-), \ \ \ \
c_{d}=\frac{{g_1'}^2}{2}({g_V^d}^2+{g_A^d}^2)Br(\ell^+\ell^-),
\label{eq:cucd}
\eeq
and $w_u(s,M_{Z'}^2)$ and $w_d(s,M_{Z'}^2)$ are related to the parton
luminosities $\left( \frac{dL_{u\overline{u}}}{dM_{Z'}^2}\right)$ and $\left(
\frac{dL_{d\overline{d}}}{dM_{Z'}^2}\right)$ and therefore only depend on the
collider energy and the $Z'$ mass. All the model dependence of the 
cross-section is therefore contained in the two coefficients, $c_u$ and $c_d$. 
These parameters can be calculated from $g_V^f, g_A^f$ and $g_1'$, assuming   
only SM decays of the $Z'$ boson. Note that the cross-section is proportional 
to $g_1'^2$ and will therefore be reduced in Little $Z'$ models in which $g_1' \ll g'\approx 0.46$.

A given model such as the E$_6$SSM \cite{King:2005jy} appears as a point in the $c_d-c_u$ plane,
assuming that the low energy $U(1)'$ gauge coupling $g_1'$ is fixed by a unification condition.
If we relax the unification condition then the point will become a line in the $c_d-c_u$ plane,
since each of $c_u$ and $c_d$ are proportional to $g_1'^2$ and the points on the line will approach the origin
as $g_1' \rightarrow 0$. For example in the E$_6$SSM we have:
\beq
c_u = 5.94\times 10^{-4}\left[ \frac{g_1'}{0.46} \right]^2, \ \ 
c_d = 1.48\times 10^{-3}\left[ \frac{g_1'}{0.46} \right]^2 .
\label{cucdgp}
\eeq
Since the experimental $Z'$ mass contours in the $c_d-c_u$ plane are fixed for a given 
limit on the cross-section, the effect of reducing $g_1'$ will not change those contours.
The only effect of reducing $g_1'$ is to move the model point in the $c_d-c_u$ plane closer to the origin,
resulting in a reduced experimental limit on the $Z'$ mass.
{See for example \cite{Accomando:2010fz} where this approach is followed 
for conventional $Z'$ models. Although this provides a simple way to understand qualitatively why the 
experimental limits are relaxed by reducing $g_1'$, it turns out that for the lower mass $Z'$ signal regions 
backgrounds and other constraints become more important for this reason
we shall not present our results in the $c_d-c_u$ plane.}

In the E$_6$SSM the $Z'$ mass is given to good approximation by:
\beq
M^2_{Z'}=g^{'2}_1 v^2\biggl(\tilde{Q}_1^2\cos^2\beta+\tilde{Q}_2^2\sin^2\beta\biggr)+g^{'2}_1\tilde{Q}^2_S s^2
\approx g^{'2}_1\tilde{Q}^2_S s^2
\,,
\label{Zpmass}
\eeq
where {the charges are }$\tilde{Q}_S=5/\sqrt{40}$,
$\tilde{Q}_1=-3/\sqrt{40}$, $\tilde{Q}_2=-2/\sqrt{40}$.
The last approximation in Eq.\ref{Zpmass} assumes
$s\gg v$ where we can neglect the terms involving the electroweak VEV $v=246$ GeV.
{What is the effect of reducing $g_1'$ in this model?}
On the one hand, reducing $g_1'$ will reduce $M_{Z'}$ in direct proportion, since
$M_{Z'}\propto g_1'$ for a fixed value of $s$.
On the other hand, reducing $g_1'$ will reduce the cross-section since $c_{u,d}\propto {g_1'}^2$
(see Eq.\ref{cucdgp}).

\begin{figure}[h!]
	\begin{center}
		\includegraphics[width=0.9\linewidth]{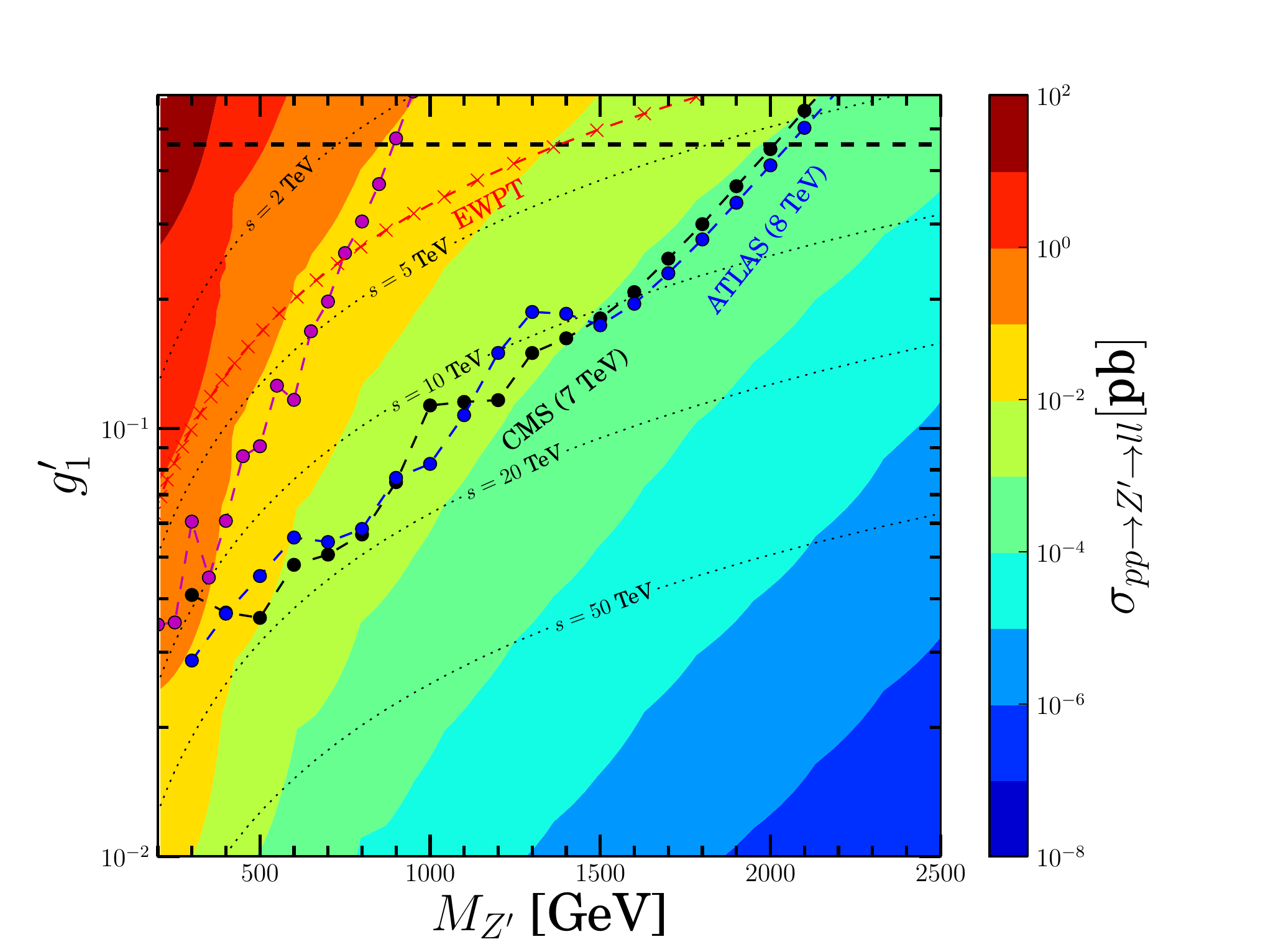}
	\end{center}
	\caption{Cross section for lepton ($e,\mu$) pair-production via $Z'$ at LHC at $\sqrt{s}= 8$ TeV in the $g_1'$ -- $M_{Z^{\prime}}$ plane{ for Little $Z'$ models with charges corresponding to the E$_6$SSM. The horizontal,} dashed line indicates the standard GUT predicted  $g_1'$ value. Exclusion limits from direct searches are plotted with dash-dotted lines in magenta, black and blue for D$\emptyset$, CMS and ATLAS respectively. Indirect constraint on the mass-coupling ratio from electro-weak precision tests are plotted with red crosses and coincides with the contour for the singlet VEV $s\approx 4$ TeV.}
	\label{fig:cs-scan-m}
\end{figure}

In Fig.~\ref{fig:cs-scan-m} we show the
cross section for lepton ($e,\mu$) pair-production via $Z'$ at LHC at $\sqrt{s}= 8$ TeV in the $g_1'$ -- $M_{Z^{\prime}}$ plane{ for the Little $Z'$ models with charges corresponding to the E$_6$SSM.
The horizontal}, dashed line indicates the standard GUT predicted  $g_1'$ value.
{
The dash-dotted lines are cross section limits on the E$_6$SSM $Z^{\prime}$ from D$\emptyset$\cite{Abazov:2010ti}, CMS\cite{:2012vb} and ATLAS\cite{ATLAS:2012ipa} that have been converted to limits on the coupling $g_1^{\prime}$. 

The estimated indirect exclusion limits from electro-weak precision tests (EWPT)\cite{Barbieri:2004qk,Cacciapaglia:2006pk,Salvioni:2009mt}, mostly by LEP, on the ratio $\frac{M_{Z^\prime}}{g_1^\prime}$ are plotted with red crosses. 
{These are not available for the E$_6$SSM but the}
limit for the $U(1)_\chi$ $Z^\prime$ is 
$$ \frac{M_{Z^\prime}}{g_1^\prime}> 3.8 \mbox{ TeV,}$$ 
and the limit for the $U(1)_\psi$ $Z^\prime$ is 
$$ \frac{M_{Z^\prime}}{g_1^\prime}> 2.5 \mbox{ TeV.}$$
{As an estimate of these limits for the E$_6$SSM,
we plot with red crosses an intermediate limit
$$ \frac{M_{Z^\prime}}{g_1^\prime}\gtrsim 3.0 \mbox{ TeV.}$$ }

Figure \ref{fig:cs-scan-m} also shows contours of constant values of the singlet VEV $s$, so it is possible to read off exclusions limits on $s$. At large masses the limits from ATLAS and CMS follow the cross section contours well but in the low mass regime the standard model background is large which weakens the limits on the cross section. In this region, just above 200 GeV, the direct searches by the LHC and Tevatron experiments place the strongest bounds on the coupling and all place limits of about $g_1^\prime < 0.03$. 

It is obvious from Fig. \ref{fig:cs-scan-m} that it is possible to lower the limit on $M_{Z^\prime}$ by decreasing the coupling $g_1^{\prime}$ but by doing this the value of the singlet VEV, $s$, generally has to increase. The limit on $s$ is however strongest for $M_{Z^\prime}$ of about 500-800 GeV and gets slightly relaxed in the lowest mass region, 200-500 GeV. Examples of how the limits on $M_{Z^\prime}$, $s$ and the fine-tuning with respect to $M_{Z^\prime}$ changes as the coupling $g_1^\prime$ decreases are tabulated in Tab.~\ref{tab:g1p}.
}

\setlength{\tabcolsep}{5.2pt}
\begin{table}
	\centering 
	\begin{tabular}{|c|c|c|c|c|c} 
		$g_1^{\prime}/0.46$& 
		1	&	1/2	&	1/10	& 1/15 	\\ \hline 
		$s>$	&	
		5.5	&	9.3	& 	16.5 	& $\sim$ 11 &  [TeV]\\
		$M_{Z^{\prime}}>$& 
		2 	& 	1.7 	& 	0.6 	& $\sim$ 0.2 &  [TeV]\\
		$\Delta_{M_{Z^{\prime}}}>$ &	
		192	&	139	& 	17	& $\sim$ 1 &	\\

	\end{tabular}
	\caption{Scenarios with different values of $g_1^{\prime}$
	{ for the Little $Z'$ models with charges corresponding to the E$_6$SSM.}
	The $Z^{\prime}$ mass and thus its source of fine-tuning, $\Delta_{M_{Z^{\prime}}}=\frac{M_{Z^{\prime}}^2}{M_Z^2}\frac{\partial M_{Z}^2}{\partial M_{Z^{\prime}}^2}$, can be reduced by reducing $g_1^\prime$ at the cost of increasing the singlet VEV, $s$. Because experimental limits on the cross section get weaker in the low mass region the limit on $s$ gets slightly weaker, hence the weaker limit on $s$ in the case of $g_1^\prime = 0.46/15$.}
	\label{tab:g1p}
\end{table}

\section{Conclusion}

{The current experimental limits from the LHC on the $Z'$ boson mass
of 2-3 TeV raises the fine-tuning in 
$E_6$ supersymmetric models to undesirably high levels.
{This is a generic property of SUSY models where the Higgs doublets carry the $U(1)^\prime$ charge.}
In order to solve this problem 
we have proposed a new class of models called {\it Little $Z'$ models} involving a weakly coupled lower mass $Z'$. These models can originate from supersymmetric $E_6$ inspired supersymmetric models 
where the spontaneously broken extra $U(1)'$ gauge group has 
a reduced gauge coupling.

We have shown that reducing the value of the extra gauge coupling relaxes these limits,
leading to the possibility of low mass $Z'$ resonances, for example down to about 200 GeV,
thereby reducing fine-tuning due to the $Z'$ mass down to acceptable levels.
Such a reduced extra gauge coupling does not affect conventional gauge coupling unification 
of the strong, weak and electromagnetic gauge couplings and in fact is well motivated in 
certain classes of F-theory models.
We emphasise the main experimental prediction of such Little $Z'$ models 
which is the appearance of a low mass weakly coupled
$Z'$ which may yet appear in future LHC searches.
Although the source of tree level fine-tuning due to the $Z'$ mass
is reduced in Little $Z'$ models, it does so 
at the expense of increasing the singlet vacuum expectation value, leading to overall fine-tuning similar to that in the Minimal Supersymmetric Standard Model.}

\section*{Acknowledgements}
SFK acknowledges partial support 
from the STFC Consolidated ST/J000396/1 and EU ITN grants UNILHC 237920 and INVISIBLES 289442 .
PS is thankful for support from the NExT institute, a part of SEPnet.

\providecommand{\bysame}{\leavevmode\hbox to3em{\hrulefill}\thinspace}

\end{document}